# Eliminating Public Knowledge Biases in Small Group Predictions

Kay-Yut Chen, Leslie R. Fine and Bernardo A. Huberman

HP Laboratories

Palo Alto, CA 94304

May 7, 2002

## Abstract

We present a novel methodology for identifying public knowledge and eliminating the biases it creates when aggregating information in small group settings. A two-stage mechanism consisting of an information market and a coordination game is used to reveal and adjust for individuals' public information. A nonlinear aggregation of their decisions then allows for the calculation of the probability of the future outcome of an uncertain event, which can then be compared to both the objective probability of its occurrence and the performance of the market as a whole. Experiments show that this nonlinear aggregation mechanism outperforms both the imperfect market and the best of the participants.

Introduction

The prediction of the future outcomes of uncertain situations is an important problem for individuals and organizations. As a result, large resources are devoted to producing reliable forecasts of technology trends, revenues, growth, and other valuable insights. To complicate matters, in the case of organizations the information relevant to predictions is often dispersed across people, making it hard to identify and aggregate it. Thus, while several methods are presently used in forecasting, ranging from committees and expert consultants to aggregation techniques such as the Delphi method [1], the results obtained suffer in terms of accuracy and ease of implementation.

In this paper, we propose and experimentally verify a market-based method to aggregate scattered information so as to produce reliable forecasts of uncertain events. This method is based on the belief shared by most economists that markets efficiently collect and disseminate information [2]. In particular, rational expectations theory tells us that markets have the capacity not only to aggregate information held by individuals, but also to convey it via the price and volume of assets associated with that information. Therefore, a possible methodology for the prediction of future outcomes is the construction of markets where the asset is information rather than a physical good. Laboratory experiments have determined that these markets do indeed have the capacity to aggregate information in this type of setting [3, 4, 5, 6].

Information markets generally involve the trading of state-contingent securities. If these markets are large enough and properly designed, they can be more accurate than other techniques for extracting diffuse information, such as surveys and opinions polls. There are problems however, with information markets, as they tend to suffer from information traps [7, 8], illiquidity [9], manipulation [10, 11], and lack of equilibrium [12, 13][1]. These problems are exacerbated when the groups involved are small and not very experienced at playing in these markets. Even when possible, proper market design is very expensive, fragile, and context-specific.

---

[1] Notable exceptions: The Iowa Electronic Market [14] has shown that political events can be accurately predicted using markets when they are large enough. Their predictions have consistently been more accurate than those resulting from major news polls. Additionally, recent work by Pennock, Lawrence, Giles and Nielsen [15] show that the Hollywood Stock Exchange (HSX) does a remarkable job of predicting box office revenues and Oscar winners. However, both of these institutions have many traders, while we focus on systems with small number of participants (fewer than 15).

In spite of these obstacles, it is worth noting that certain participants in information markets can have either superior knowledge of the information being sought, or are better processors of the knowledge harnessed by the information market itself. By keeping track of the profits and final holdings of the members, one can determine which participants have these talents, along with their risk attitudes.

In earlier work, (Chen, Fine, and Huberman [16]), we demonstrated the comparative efficacy of a nonlinear aggregation mechanism with behavioral components to that of a market. Specifically, we showed that one could take past predictive performance of participants in information markets and to create weighting schemes that help predict future events, even if they are not the same event on which the performance was measured. Furthermore, our two-stage approach successfully harnessed distributed knowledge in a manner that alleviated the problems that arise from low levels of participation.

However, these results were not robust to the presence of public information, that is, information that is commonly known to multiple individuals in the group. This is because public information is bound to introduce strong correlations in the knowledge possessed by members of the group, correlations that were not explicitly taken into account by our aggregation algorithm.

Nevertheless, the success of our two-stage forecasting mechanism with private information led us to search for suitable modifications that would allow the detection of the amount of public information present in a group so as to subtract it. Assuming that subjects can differentiate between the public and private information they hold, that the private aspect of their information is truly private (held only by one individual), and that the public information is truly public (held by at least two individuals), we create a coordination variant of the mechanism which allows for the identification of public information within a group and its subtraction when aggregating individual predictions about uncertain outcomes. Experiments in the laboratory show that this aggregation mechanism outperforms both the market and the best player in the group.

In what follows, we first outline the original two-stage mechanism for information aggregation and then explain the modified second stage that allows for public information to be extracted. Next, we present laboratory experiments that quantitatively measure the performance of this new mechanism and established its superiority with respect to both the information market and the participating members. An appendix provides the mathematical details of the coordination game.

### Extracting Private Information

We start by reviewing the original nonlinear aggregation scheme, presented in Chen, Fine and Huberman [16]. This aggregation scheme applies to a group of individuals that hold private information.

In the first stage of this mechanism, individuals participate in an information market designed to elicit their risk attitudes and other relevant behavioral information. As was shown, although the participant pool is too small for the market to act perfectly efficiently, it nevertheless provides accurate behavioral information. In the second stage, individuals are asked to report their beliefs and these beliefs and are in turn aggregated in a nonlinear fashion that takes into account the behavioral information gathered in the first stage. The two stages are applied to different events that are nevertheless structurally similar.

In the second stage, each player is asked to report a vector of perceived state-probabilities, $\{p_1, p_2, \ldots p_N\}$, with $p_i$ the probability that a given state $i$ will be realized, and with the constraint that the vector sums to one. When the true state $x$ is revealed, each player is paid an amount equal to $c_1 + c_2 * log(p_x)$, where $c_1$ and $c_2$ are positive numbers. This payoff function ensures that risk-neutral expected utility maximizers would report their true beliefs. So, each player should report his perceived probability distribution over the $N$ possible states.

In order to compute the probability distribution, we aggregate the individual reports by using the following nonlinear aggregation function, which is a modification of Bayes' rule:

$$P(s \mid I) = \frac{p_{s_1}^{\beta_1} p_{s_2}^{\beta_2} \cdots p_{s_N}^{\beta_N}}{\sum_{\forall s} p_{s_1}^{\beta_1} p_{s_2}^{\beta_2} \cdots p_{s_N}^{\beta_N}} \quad (2)$$

where *s* is a given possible state, *I* is the available information, and $\beta_i$ is the exponent assigned to individual *i*. The role of $\beta_i$ is to help recover the true posterior probabilities from individual *i*'s report. The value of $\beta$ for a risk neutral individual is one, as he should report the true probabilities coming out of his information. For a risk averse individual, $\beta_i$ is greater than one so as to compensate for the flat distribution that he reports. The reverse, namely $\beta_i$ smaller than one, applies to risk loving individuals. In terms of both the market performance and the individual holdings and risk behavior, $\beta_i$ is given by

$$\beta_i = r(V_i / \sigma_i) c \quad (3)$$

where *r* is a parameter that captures the risk attitude of the whole market and is reflected in the market prices of the assets, $V_i$ is the utility of individual *i*, and $\beta_i$ is the variance of his holdings over time. We use *c* as a normalization factor so that if *r*=1, $\sum \beta_i$ equals the number of players.

### Identifying Public Information

Although this mechanism works well with private independent information (see Tables 1 and 2, Experiments 1 through 5), its performance can be significantly degraded by the introduction of public information. The introduction of public information implies that the probabilities that enter into Equation (2) are no longer independent of each other, and therefore they are no longer aggregated correctly. Equation (2) over counts information that is observed by more than one individual since it adds (in the probability space) probabilities disregarding whether the reports are coming from the same information source.

Thus the mechanism has to incorporate a feature that distinguishes the public information from the private, so that it can be suitably subtracted when aggregating the individual predictions. We achieve this by using in the second stage a coordination game, which incents players to reveal what they believe others will reveal. This coordination game is similar to the Battle of the Sexes game.

In the Battle of the Sexes, a couple enjoys spending time together, but each member would rather do so while engaged in his or her preferred activity. As an example, a payoff matrix is shown for an instance in which he'd like them both to go to the baseball game (upper-left), and she'd prefer they went to the opera together (lower-right). If they disagree, no one goes anywhere and no one is happy (off-diagonals).

|  |  | SHE | |
|---|---|---|---|
|  |  | B | O |
| HE | B | 3,1 | 0,0 |
|  | O | 0,0 | 1,3 |

This game has multiple mixed-strategy Nash equilibria, in which both players mix with the goal of landing in the upper-left and lower-right quadrants of the payoff matrix. Notice that with these payoffs each member of the couple is incented to reveal the information that they believe the other will.

In much the same way, our matching game asks players that, in addition to making their best bet (MYBB), they reveal what they believe they all know (AK). The first half, MYBB, works as in the original experiments. That is, players report a vector of bets on the possible states, and are paid according to a log function of these bets. In the AK game however, the subjects try to guess the bets placed by *someone else* in the room, and these bets are then matched to another player whose bets are most similar to theirs. The payout from this part of the game is a function of both their matching level and the possible payout from the number of tickets allocated by the other member of the pair. The payoffs are constructed such that participants have the incentive to match their peers in their public reports. The design of this game is discussed further in the Experimental Design section.

In order to design a payoff function that induces both truthful revelation and maximal matching, we assume that: (A1) the public and private information held by an individual are independent of one another, (A2) that private information is independent across individuals, (A3) that public information is truly public (observed by more than one individual), and (A4) that an given individual can distinguish between the public and the private information he holds. In other words:

For each individual $i$ with observed information $O_i$, there exists information $O_i^{priv}$ and $O_i^{pub}$ such that:

(A1) $P(s|O_i) = P(s|O_i^{priv}) P(s|O_i^{pub})$ for all $i,s$

(A2) $P(s|O_i^{priv}$ and $O_j^{priv}) = P(s|O_i^{priv}) P(s|O_j^{priv})$ for all $i,j,s$

(A3) There exists a $j$ for every $i$ such that $O_i^{pub} = O_j^{pub}$

(A4) All individuals know $P(s|O_i^{priv})$ and $P(s|O_i^{pub})$

So, in the second stage, each player $i$ is asked to report *two* probability distributions, $\vec{p}_i = \{p_{i1}, p_{i2}, ... p_{iN}\}$ (from MYBB) and $\vec{q}_i = \{q_{i1}, q_{i2}, ... q_{iN}\}$ (from AK), by allocating a set of tickets to each of the possible states. Let $x$ be the true outcome. The payoff function for each player $i$ is given by the following expression:

$$P = c_1 + c_2 * \log(p_{ix}) + f(\vec{q}_i, \vec{q}_j) * (c_4 + c_5 * \log(q_{jx})) \quad (4)$$

where $c_1$, $c_2$, $c_3$ and $c_4$ are positive constants, $j$ is chosen in such a way that $f(\vec{q}_i, \vec{q}_j) \geq f(\vec{q}_i, \vec{q}_k)$ for all $k$, and the function $f(.)$ is given by:

$$f(\vec{x}, \vec{y}) = \left(1 - \left[\sum_s |x_s - y_s|\right]/2\right)^2 \vec{y} \quad (5)$$

In words, subjects are paid according to a log function of their reports in the MYBB game, plus a payment from the AK game. This payment is a function of the player with whom he has a maximal match, and is the product of the matching level and a scaled log function of the *matched* player's report in the AK game. This match level is given by the second term of Equation (4) and is detailed in Equation (5) above.

As shown earlier (Chen, Fine and Huberman [16]), the first part of the payoff function in Equation (4), $c_1 + c_2 * \log(p_{ix})$, will induce risk neutral subjects who maximize their expected utility to report their true belief, conditioned on *both* their private and public information. Concerning the last term of Equation (4), we first note that player *i* can only affect it through his matching level, which is given by the function $f(\vec{q}_i, \vec{q}_j)$. Since $f(\vec{x},\vec{x}) \geq f(\vec{y},\vec{x})$ for all $\vec{y}$, player *i*'s best response is to report $\vec{q}_i = \vec{q}_j$. Further, since *j* is chosen such that $f(\vec{q}_i, \vec{q}_j) \geq f(\vec{q}_i, \vec{q}_k)$ for all *k*, player *i* only needs to co-ordinate his $\vec{q}_i$ with only one other individual in the group to achieve an optimal payoff. Additionally, it is easy to show that this part of the game has multiple Nash equilibria, since any common report vector $\vec{q}$ reported by both players *i* and *j* is a potential Nash equilibrium. Therefore, we designed the payoff function given by in Equations (4) and (5) to encourage individuals to coordinate on the probability distribution induced by the public information. Lastly, the third piece of the payoff function for player *i*, $c_4 + c_5 * log(q_{jx})$ induces a different payoff for each Nash equilibrium $\vec{q}$ on which the two individuals coordinate. Since this factor depends on the strategy of player *i*'s partner *j*, no one player can directly affect it. This is crucial to preserve the equilibrium structure.

We thus designed the payoff such that the more information revealed in the reports $\vec{q}$, the higher the potential payoff to the subjects involved, which implies an information-rich equilibrium. Additionally, since private information is independent across individuals (it is truly private), the best equilibrium on which individuals can coordinate on is the probability distribution induced by using the public information only. Therefore, this mechanism will induce individuals to report both their true beliefs ($\vec{p}_i$) and their public information ($\vec{q}_i$). Once these vectors are reported, we still need to aggregate them, which we discuss in the next section.

### Aggregating Information

Once we have a mechanism for extracting public beliefs from private ones, it is straightforward to add a public information generalization to Equation (2). By dividing the perceived probability distributions of the players by the distributions induced by the public information only, we develop what we call a *General Public Information Mechanism* (GPIC), which is given by

Page 8

$$P(s \mid I) = \frac{\left(\frac{p_{s1}}{q_{s1}}\right)^{\beta_1} \left(\frac{p_{s2}}{q_{s2}}\right)^{\beta_2} \cdots \left(\frac{p_{sN}}{q_{sN}}\right)^{\beta_N}}{\sum_{\forall s} \left(\frac{p_{s1}}{q_{s1}}\right)^{\beta_1} \left(\frac{p_{s2}}{q_{s2}}\right)^{\beta_2} \cdots \left(\frac{p_{sN}}{q_{sN}}\right)^{\beta_N}} \quad (6)$$

where the *q*s are extracted from individuals' reports before they are aggregated. This correction allows us to isolate the private information from the individual reports.

While this mechanism is quite general, and outperforms both the market prediction and that of our original IAM, there are potential improvements to it that can be implemented. Thus, we developed modifications to the aggregation function to address issues of uncertain information structures and multiple equilibria. In theory, knowledge of the individuals' reports $\vec{p}_i = \{p_{i1}, p_{i2}, \ldots p_{iN}\}$ and $\vec{q}_i = \{q_{i1}, q_{i2}, \ldots q_{iN}\}$, should make information aggregation straightforward since for a given individual *i*, his probability assignment to state *s*, with respect to private information, should be proportional to $p_{si}/q_{si}$. To more efficiently add in public information, we aggregate the individual reports of public information $\vec{q}_i = \{q_{i1}, q_{i2}, \ldots q_{iN}\}$ into a single vector $\vec{q} = \{q_1, q_2, \ldots q_N\}$. In order to do this, we employ one additional assumption, that every individual observes the *same* public information, $O^{pub}$. We then aggregate by averaging the reports, weighted by each individual's $\beta$, thusly:

$$q_s = \sum_{i=1}^{N} \beta_i q_{si} \Big/ \sum_{i=1}^{N} \beta_i \quad (7)$$

Once we have completed this aggregation process, we can use the new vector $\vec{q}$ in place of $\vec{q}_i$ in the original function in Equation (6). If $\vec{q}$ is derived correctly, it will resolve the matter of parsing the private information from the public. Furthermore, in much the same way that some people process their private signals better than others, there are some individuals that report public information more accurately than others. If one can identify these individuals, one can recover public information more efficiently than by taking a weighted average of everyone's report. Thus, instead of using the whole group to recover public information, as in Equation (7), we use a limited set J, a subset of the whole group:

$$q_s = \sum_{i \in J} \beta_i q_{si} \bigg/ \sum_{i=1}^{N} \beta_i \qquad (7a)$$

The resultant forecast is then determined by a modification of the GPIC in Equation (6). It uses a small subset of players to determine the public information so as to parse it from the private. While this mechanism is quite efficient, it only applies to the special case where the public information is completely public and identical. Therefore, we refer to it as the *Special Public Information Correction Mechanism*, or SPIC.

$$P(s \mid I) = \frac{q_s \left(\frac{p_{s1}}{q_s}\right)^{\beta_1} \left(\frac{p_{s2}}{q_s}\right)^{\beta_2} \cdots \left(\frac{p_{sN}}{q_s}\right)^{\beta_N}}{\sum_{\forall s} q_s \left(\frac{p_{s1}}{q_s}\right)^{\beta_1} \left(\frac{p_{s2}}{q_s}\right)^{\beta_2} \cdots \left(\frac{p_{sN}}{q_s}\right)^{\beta_N}} \qquad (8)$$

### Experimental Design

In order to test this mechanism we conducted a number of experiments at Hewlett-Packard Laboratories in Palo Alto, California. The subjects were undergraduate and graduate students at Stanford University and knew the experimental parameters discussed below, as they were part of the instructions and training for the sessions. The five sessions were run with nine to eleven subjects in each.

We implemented the two-stage mechanism in a laboratory setting. Possible outcomes were referred to as states in the experiments. There were ten possible states, A through J, in all the experiments. Each had an Arrow-Debreu state security associated with it. The information available to the subjects consisted of observed sets of random draws from an urn with replacement. After privately drawing the state for the ensuing period, we filled the urn with one ball for each state, plus an additional two balls for the true state security. Thus it is slightly more likely to observe a ball for the true state than others.

We allowed subjects to observe different number of draws from the urn in order to controlled the amount of information given to the subjects. We used four variants on the of information structure to ensure that the results obtained were robust. In the first structure, we provide two private and two public draws to all participants. In the remaining, all subjects received three private draws. In one, they also received one public draw, in another, only half of the cohort received the public draw, and in the final treatment all players received a public draw, but there were two *different* public draws available. Further details of the treatments can be found in Table 1.

The information market we constructed consists of an artificial call market in which the securities are traded. The states are equally likely and randomly drawn. If a state occurred, the associated state security paid off at a value of 1,000 francs. Hence, the expected value of any given security, a priori, is 100 francs. Subjects were provided with some securities and francs at the beginning of each period.

Each period consisted of six rounds, lasting 90 seconds each. At the end of each round, the bids and asks were gathered and a market price and volume was determined. The transactions were then completed and another call round began. At the end of six trading rounds the period was over, the true state security was revealed, and subjects were paid according to the holdings of that security. This procedure was then repeated in the next period, with no correlation between the states drawn in each period.

In the second-stage, every subject played under the same information *structure* as in the first stage, although the draws and the true states were independent from those in the first. There are two parts to this game, described in the Identifying Public Information section above, which were referred to as the "What Do We All Know" (AK) and the "Make Your Best Bet" (MYBB) games. Each period, the subjects received their draws of information, as in the market game. They also received two sets of 100 tickets each, one set for AK, and one for MYBB. We will discuss these two games in turn.

In MYBB, the subjects were asked to distribute their tickets across the ten states with the constraint that all 100 tickets must be spent each period and that at least one ticket is spent on each state. Since the fraction of tickets spent determines $p_{si}$, this implies that $p_{si}$ is never zero. The subjects were given a chart that told them how many francs they would earn upon the realization of the true state as a function of the number of tickets spent on the true state security. The payoff was a linear function of the log of the percentage of tickets placed in the winning state as given by the first half of Equation (4). The chart the subjects received showed the payoff for every possible ticket expenditure, and an excerpt from the chart is shown in Table 3. The MYBB game is identical to the second stage played in Chen, Fine and Huberman [16].

We also played the matching game in this stage, known as AK. In this stage, subjects received 100 tickets, but with a different goal. They tried to guess the bets placed by someone else in the room. After they placed the bets, they were matched to another player, one whose bets were most similar to theirs. The more similar the bets were to their nearest match, the higher the reported "Percent Match with Partner." The payoffs for any given ticket expenditure were higher in the AK game than the MYBB game, and are detailed in Table 3.

Figure 1 shows a screenshot from the second stage of the game, which displays the bets placed in a sample Period 1. As shown on the upper right, the true state was F. Following down the items reported in the upper right of the screen, we see that this player bet 20 tickets on F in the MYBB game, which has corresponds to a Possible Payout of 662 francs. He was matched with a partner whose AK distribution of tickets matched his at a 49% level. This partner bet enough tickets to have a Possible Payout of 178 francs. Our sample player thus earned 662 francs for the 20 tickets bet in the MYBB game, plus .49*178 = 87 francs for the AK game, for a total of 749 francs.

Figure 1: Sample Page from Stage Two of the Experiment

| Possible States: | A | To | J | Period | | 1 |
|---|---|---|---|---|---|---|
| Drawing from a urn with replacement containing: | | | | State | | F |
| 3 | balls for the true state | | Total | Tickets on this State | | 20 |
| 1 | ball for each false state | | Payoff | Percent Match with Partner | | 49 |
| \|\|\| | Private Information | | 749 | Maximal Payoff from Partner | | 178 |
| >>> | Public Information | | | Payoff | | 749 |
| | MAKE YOUR BEST BET | | WHAT DO WE ALL KNOW? | My Information | | |
| State | Number of Tickets | Possible Payout | Number of Tickets | | State | Count | |
| A | 20 | 662 | 10 | | A | 1 | \|\|\| |
| B | 3 | 264 | 5 | | B | 0 | |
| C | 3 | 264 | 5 | | C | 0 | |
| D | 3 | 264 | 5 | | D | 0 | |
| E | 20 | 662 | 25 | | E | 1 | >>> |
| F | 20 | 662 | 10 | | F | 1 | \|\|\| |
| G | 3 | 264 | 5 | | G | 0 | |
| H | 3 | 264 | 5 | | H | 0 | |
| I | 20 | 662 | 25 | | I | 1 | >>> |
| J | 5 | 371 | 5 | | J | 0 | |
| | Total Tickets Spent | 100 | Total Tickets Spent | 100 | | | |

Analysis

In order to analyze these results we first calculate an omniscient probability distribution for each period using every observation that was available to the individuals. This distribution is used as a limit-case benchmark. That is, only a perfect information aggregation mechanism should be able to achieve this distribution. We compare the resultant probabilities from information aggregation mechanisms to this benchmark by using the Kullback-Leibler measure [17]. The Kullback-Leibler measure of two probability distributions $p$ and $q$ is given by:

$$KL(p,q) = E_p\left(\log\left(\frac{p}{q}\right)\right) \quad (9)$$

where $p$ is the "true" distribution (in our case, the omniscient probability distribution). In the case of finite number of discrete states, the above Equation (9) can be rewritten as:

$$KL(p,q) = \sum_s p_s \log\left(\frac{p_s}{q_s}\right) \quad (10)$$

It can be shown that $KL(p,q)=0$ if and only if the two distributions being compared, $p$ and $q$, are identical, and that $KL(p,q) \geq 0$ for all probability distributions $p$ and $q$. Therefore, the smaller the Kullback-Leibler number, the closer that two probabilities are to each other. Furthermore, the Kullback-Leibler measure of the joint distribution of multiple independent events is the sum of the Kullback-Leibler measures of the individual events. Since periods within an experiment were independent events, the sum or average (across periods) of Kullback-Leibler measures is a good summary statistic for an entire experiment.

We compare five information aggregation mechanisms to the benchmark distributions. In addition, we also report the Kullback-Leibler measures of the no information prediction (uniform distribution over all the possible states) and the best (most accurate) individual's predictions. The no information prediction serves as the first baseline to determine if any information is contained in our mechanism's predictions. Further, if a mechanism is really aggregating information, then it should be doing at least as well as the best individual. Therefore, the predictions of the best individual in the experiment serve as the second baseline, which helps us to determine if information aggregation indeed occurred in the experiments.

The first two information aggregation mechanisms we evaluate are the market prediction and the Chen, Fine, and Huberman [16] mechanism in Equation (2). We calculate the market prediction by using the last traded prices of the assets. We use the last traded prices rather than the current round's price because sometimes there was no trade in a given asset in a given round. From these prices, we infer a probability distribution on the states. The second aggregation mechanism is the original IAM, found in Equation (2). Recall that this mechanism was designed on the assumption of no public information. The purpose of its inclusion is to measure the performance degradation due to the double-counting issue of inherent to the presence of public information.

The third mechanism is our proposed improvement, referred to as the *General Public Information Correction* (GPIC) *mechanism*, given by Equation (6). It uses both individuals' reports of public information regarding outcomes as well as the individuals' perceived probabilities of these outcomes. If this mechanism is working as predicted by the theory, it should provide a superior outcome to that of the original IAM.

As an additional benchmark the fourth mechanism, referred to as the *Perfect Public Info Correction* (PPIC), replaces individuals' reports of public information with the true public information that they have observed. Obviously, this is not possible in a realistic environment, since we do not know the true public information (or, this exercise would be pointless). However, it allows us to validate the behavioral assumptions we make in the design of the mechanism. Our model implicitly assumes that individuals aggregate their public and private information by a modified version of Bayes' rule to arrive at their reports, and we can use this benchmark to validate this assumption.

Lastly, we address the special case in which the experimenter knows that every individual receives the same public information. This fifth mechanism, referred to as the *Special Public Info Correction mechanism* (SPIC), recovers the public information by using the reports of only the best two individuals to correct the public information bias in all participants' reports.

Results

We start by reporting not only the result of our public information experiments, but those of the original Chen, Fine, and Huberman [16] paper as well (Experiments 1 through 5 on Tables 1 and 2). Recall that in these experiments, all information was independent and private. As is shown in Tables 1 and 2, once even a small amount of public information is introduced into the system (Experiments 6 through 10), the performance of the original IAM decreases dramatically. In Figure 2 we illustrate the double counting issue before the GPIC modification. In this figure, we plot the probability distributions generated by omniscience, the prediction from the original

IAM and the available public information from a sample period (Experiment 8, period 9). As one can see, using the original IAM results in a false peak at state H, which is the state on which public information was available. In some cases, the double counting issue is so severe that the results are worse than that of the no information measure (see, for example, Experiments 6, 7 and 8). Thus, this verifies the necessity to derive a method correcting for the biases introduced by public information.

Figure 2: Illustration of the Double Counting Issue

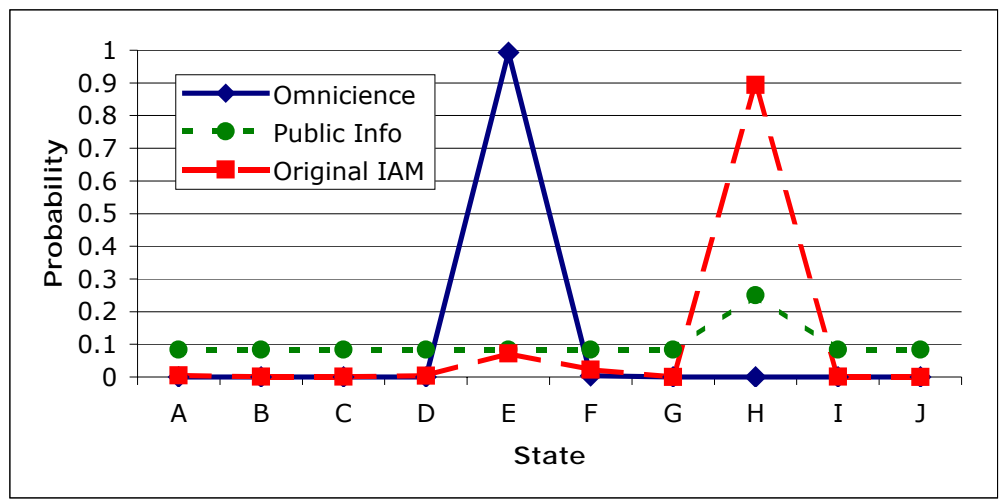

In Table 1 we summarizes the relative performance, in terms Kullback-Leibler measures, of all of the benchmarks mechanisms enumerated above. Table 2 reports the same results in terms of the percentage relative to the no information Kullback-Leibler measure (indicating the level of improvement over this benchmark). Note that the amount of aggregate information available in an experiment varied across the treatments. Because the pure KL measure reported in Table 1 is affected by the amount of underlying information, the percentage measurement in Table 2 are more useful when comparing results across experiments.

The GPIC mechanism (Equation 6) outperforms the best single individual's guesses reports in all five experiments. It also outperforms the market prediction in four out of five experiments. The GPIC mechanism uses the reports of public information of individuals to perform the correction. As expected, this mechanism recovers enough public information to perform well compared to an information market. However, there is room for improvement compared to the case where the true public information is used.

To understand this inefficiency, let us assume that the information aggregator knows the true public information seen by every individual and applies the algorithm in Equation (8). The accuracy of the results obtained (Perfect Public Info Correction, or PPIC) are almost as good as the performance of the original IAM mechanism in the private information case (Experiments 1 through 5). Furthermore, this method outperforms any other method by a large margin. Although this is not an implementable mechanism, since no one knows the true public information, it does show the correctness of our behavioral model as to how people mix private and public information is correct. Therefore, there is validity in our approach to teasing out this public information in the GPIC.

Figure 3 illustrates the efficacy of the GPIC. In this figure, once again, the results from Experiment 8, period 9 are plotted. The GPIC mechanism eliminates the false peak shown in Figure 2. However, the correction is not perfect. There is still some residual positive probability being placed on state H, the site of the false peak. When the PPIC is used to perform the correction, the false peak is completely eliminated.

Figure 3: Information Aggregation with Public Information Correction

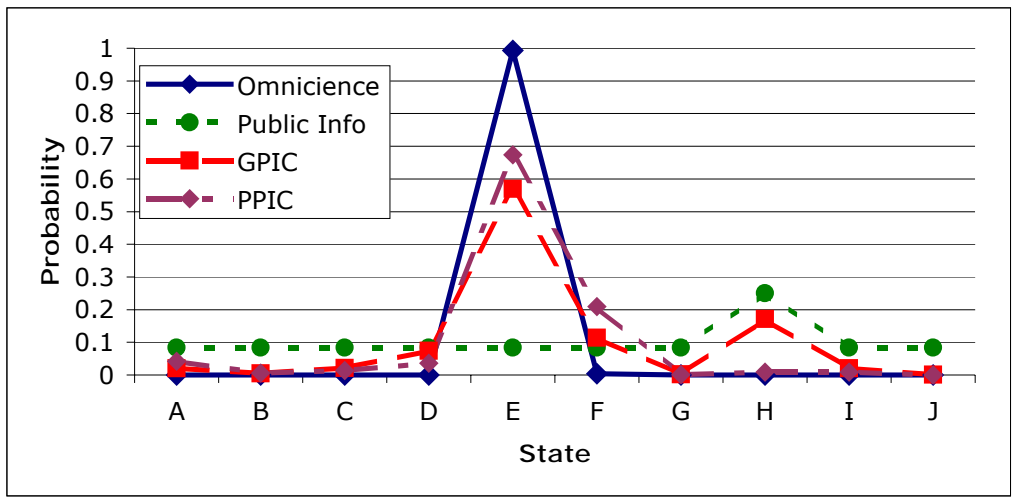

It is important to realize that while algorithms that explicitly aggregate private and public information are sensitive to the underlying information structures, markets are not. In all the experiments, including the ones with only private information, the performance of the market, measured as a percentage of the no information KL, is fairly consistent, albeit somewhat inaccurate.

It is interesting to note that if we assume that every individual receives the same public information, we may not need to use everyone's report to recover public information, as described in the SPIC mechanism By searching for pairs with the best performance, we can achieve improvements over our GPIC. However, these pairs were found ex post. That is, we calculate the performance for every pair and then choose the best. So, this extension shows merely the possibility of using pairs (or larger subgroups) to recover public information. Simple intuitive ad hoc rules, such as choosing the pairs that are closest together in the KL sense, can find good pairs in some experiments. We include the results from such an attempt in Tables 1 and 2 as the *Special Public Information Correction*, or SPIC. The issue of identifying subgroups to recover either public, or for that matter, private information is subject of future research.

## Conclusion and Extensions

Accurate predictions are essential to individuals and organizations, and are quickly becoming a key differentiator in today's economy. For large communities, information relevant to forecasts is often dispersed across people, frequently in different geographical areas. Our methodology addresses the needs for an implementable mechanism to aggregate this information accurately and with the correct incentives. One can take past predictive performance of participants in information markets and create weighting schemes that will help predict future events, even if they are not the same event on which the performance was measured. Furthermore, our two-stage approach can improve upon predictions by harnessing distributed knowledge in a manner that alleviates problems with low levels of participation. It also mitigates the issues of redundant, public signals in a group.

The rapid advances of information technologies and the understanding of information economics have opened up many new possibilities for applying mechanism design to gather and analyze information.  This paper discusses one such design and provides empirical evidence about its validity.

Table 1: Kullback-Leibler Numbers, by Experiment

| | Experimental Structure | | | Kullback-Leibler Values (Standard Deviation) | | | | | | |
|---|---|---|---|---|---|---|---|---|---|---|
| Expt | Number of Players | Private Info | Public Info | No Info | Market Prediction | Best Player | Original IAM | General Public Info Correction | Perfect Public Info Correction | Special Public Info Correction |
| 1 | 13 | 3 draws for all | None | 1.977 (0.312) | 1.222 (0.650) | 0.844 (0.599) | 0.553 (1.057) | N/A | N/A | N/A |
| 2 | 9 | 3 draws for all | None | 1.501 (0.618) | 1.112 (0.594) | 1.128 (0.389) | 0.214 (0.195) | N/A | N/A | N/A |
| 3 | 11 | ½: 5 draws ½: 1 draw | None | 1.689 (0.576) | 1.053 (1.083) | 0.876 (0.646) | 0.414 (0.404) | N/A | N/A | N/A |
| 4 | 8 | ½: 5 draws ½: 1 draw | None | 1.635 (0.570) | 1.136 (0.193) | 1.074 (0.462) | 0.413 (0.260) | N/A | N/A | N/A |
| 5 | 10 | ½: 3 draws ½: varied draws | None | 1.640 (0.598) | 1.371 (0.661) | 1.164 (0.944) | 0.395 (0.407) | N/A | N/A | N/A |
| 6 | 10 | 2 draws for all | 2 draws for all | 1.332 (0.595) | 0.847 (0.312) | 0.932 (0.566) | 2.095 (.196) | 0.825 (0.549) | 0.279 (0.254) | 0.327 (0.247) |
| 7 | 9 | 2 draws for all | 2 draws for all | 1.420 (0.424) | 0.979 (0.573) | 0.919 (0.481) | 2.911 (2.776) | 0.798 (0.532) | 0.258 (0.212) | 0.463 (0.492) |
| 8 | 11 | 3 draws for all | 1 draws for all | 1.668 (0.554) | 1.349 (0.348) | 1.033 (0.612) | 2.531 (1.920) | 0.718 (0.817) | 0.366 (0.455) | 0.669 (0.682) |
| 9 | 10 | 3 draws for all | ½: 1 draw | 1.596 (0.603) | 0.851 (0.324) | 1.072 (0.604) | 0.951 (1.049) | 0.798 (0.580) | 0.704 (0.691) | 0.793 (0.706) |
| 10 | 10 | 3 draws for all | 1 draws for all 2 sets of public info | 1.528 (0.600) | 0.798 (0.451) | 1.174 (0.652) | 0.886 (0.763) | 1.015 (0.751) | 0.472 (0.397) | 0.770 (0.638) |

Table 2: Percentage of No-Info Kullback-Leibler Numbers, by Experiment

| | Experimental Structure | | | Kullback-Leibler Values, as a Percent of the No Info Case | | | | | | |
|---|---|---|---|---|---|---|---|---|---|---|
| Expt | Number of Players | Private Info | Public Info | No Info | Market Prediction | Best Player | Original IAM | General Public Info Correction | Perfect Public Info Correction | Special Public Info Correction |
| 1 | 13 | 3 draws for all | None | 100% | 61.8% | 42.7% | 28.0% | N/A | N/A | N/A |
| 2 | 9 | 3 draws for all | None | 100% | 74.1% | 75.2% | 14.3% | N/A | N/A | N/A |
| 3 | 11 | ½: 5 draws ½: 1 draw | None | 100% | 62.3% | 51.9% | 24.5% | N/A | N/A | N/A |
| 4 | 8 | ½: 5 draws ½: 1 draw | None | 100% | 69.5% | 65.7% | 25.3% | N/A | N/A | N/A |
| 5 | 10 | ½: 3 draws ½: varied draws | None | 100% | 83.6% | 71.0% | 24.1% | N/A | N/A | N/A |
| 6 | 10 | 2 draws for all | 2 draws for all | 100% | 63.6% | 70.0% | 157.3% | 61.94% | 20.94% | 24.53% |
| 7 | 9 | 2 draws for all | 2 draws for all | 100% | 69.0% | 64.7% | 205.0% | 56.2% | 18.2% | 32.6% |
| 8 | 11 | 3 draws | 1 draws for | 100% | 80.9% | 61.9% | 151.7% | 43.0% | 22.0% | 40.1% |

| | | for all | all | | | | | | | |
|---|---|---|---|---|---|---|---|---|---|---|
| 9 | 10 | 3 draws for all | ½: 1 draw | 100% | 53.3% | 67.1% | 59.6% | 50.0% | 44.1% | 49.7% |
| 10 | 10 | 3 draws for all | 1 draws for all 2 sets of public info | 100% | 52.2% | 76.9% | 57.9% | 66.4% | 30.9% | 50.4% |

Table 3: Excerpt from Payoff Chart used in the MYBB Game

| Number of Tickets | Possible Payoff in MYBB Game | Possible Payoff in AK Game | Number of Tickets | Possible Payoff in MYBB Game | Possible Payoff in AK Game |
|---|---|---|---|---|---|
| 1 | 33 | -1244 | 50 | 854 | 1515 |
| 10 | 516 | 388 | 60 | 893 | 1642 |
| 20 | 662 | 873 | 70 | 925 | 1750 |
| 30 | 747 | 1157 | 80 | 953 | 1844 |
| 40 | 808 | 1359 | 90 | 978 | 1926 |

Appendix I

Consider the following scenario:

- $N$ possible states of the world
- $M$ players indexed by $i=1...M$
- Player $i$ is given information about the state of the world $x \in \{1,2,...,N\}$
  - His beliefs as to the probabilities of the states of the world conditioned on his information are $P_{ix}$
  - Some of player $i$'s information is observed by at least one other player $j$. Let $Q_{ix}$ be the probability conditioned on $i$'s public information only (does not consider his private information).
- Each player $i$ is asked to report two probability distributions $p_i=\{p_{i1}, p_{i2},...p_{iN}\}$ and $q_i = \{q_{i1}, q_{i2},...q_{iN}\}$ with the constraints $\sum_{s=1}^{N} p_{is} = 1$ and $\sum_{s=1}^{N} q_{is} = 1$
- The true state $x$ is revealed and he is paid $f(p_i, q_i, q_{-i}/x)$.

Assumptions:

a) Players are risk neutral utility maximizers, and
b) $f(p_i, q_i, q_{-i}/x) = c_1 + c_2 \log(p_{ix}) + H(g(q_i, q_j), \log(q_{jx}))$
c) $g: q \times q \rightarrow \Re$ is any real function of two probability distribution such that
   $$y = \underset{x}{Max}\, g(x, y)$$
d) $H(x,y)$ is increasing both in $x$ and $y$
e) $j$ is determined by: $j = \underset{k \in \{1...M\}}{\arg\max}\, g(q_i, q_k)$

*Lemma 1:* $\{p_i = P_i, q_i = Q_i \text{ for all } i\}$[2] is a Bayesian Nash equilibrium. That is, each player will report his true conditional probability beliefs and the beliefs conditioned solely on his public information.

Proof:

Assuming all players but $i$ are playing an equilibrium strategy, player $i$'s maximization problem is

$$\underset{\{p_i q_i\}}{Max} \sum_{s=1}^{N} P_{is} \{c_1 + c_2 \log(p_{is}) + H(g(q_i, Q_i), \log(Q_{is}))\} \text{ s.t. } \sum_{s=1}^{N} p_{is} = 1 \text{ and } \sum_{s=1}^{N} q_{is} = 1.$$

There will be at least one other player $j$ that plays $q_j = Q_{ix}$ since at least one player other than $i$ observes the same public information and arrives at the same distribution $Q_i$.

The resultant Langrangian is

---

[2] Notice that $p_i$ and $q_i$ are probability distributions. Therefore, the statement is equivalent to $\{p_{ix} = P_{ix}$ for all $i, x$ ; $q_{ix} = Q_{ix}$ for all $i, x\}$.

$$L = \sum_{s=1}^{N} P_{is}\{c_1 + c_2 \log(p_{is}) + H(g(q_i, Q_i), \log(Q_{is}))\} - \lambda\left(\sum_{i=1}^{N} p_{is} - 1\right) - \mu\left(\sum_{i=1}^{N} q_{is} - 1\right)$$

The first order condition is $\dfrac{P_{is}}{p_{is}} = \lambda$ for all $i$ => $P_{is} = \lambda p_{is}$

Summing over both sides, we get $1 = \lambda$. Thus $p_{is} = P_{is}$ for all $i$.

Recalling assumption (c), $q_i = Q_i$ maximizes $g(q_i, Q_i)$. Since $H$ is increasing in $g$, it also maximizes $H(g(q_i, q_j), \log(q_{jx}))$. QED.

*Lemma 2:* There are multiple equilibria to this game.

The same proof applies to $\{p_i = P_i$ for all $i$; $q_{ix} = \dfrac{1}{N}$ for all $i, x\}$ or for that matter, any set of $q_i$ on which players coordinate.